\documentclass[aps,prl,reprint, superscriptaddress,twocolumn,nobibnotes,]{revtex4-2}  
\usepackage[pdftex]{graphicx}
\usepackage[pdftex]{epsfig}
\usepackage{amsmath}
\usepackage{amssymb}
\usepackage{amsfonts}
\usepackage{color}
\usepackage{wrapfig}
\usepackage{eucal}
\usepackage{hhline}
\usepackage{threeparttable}
\usepackage{supertabular}
\usepackage{multirow}
\usepackage{tabularx}
\usepackage{siunitx}
\usepackage{upgreek}
\usepackage{float}

\usepackage[export]{adjustbox}

\usepackage{soul}

\usepackage[version=4]{mhchem} 
\usepackage{array} 
\usepackage[colorlinks=true, allcolors=blue]{hyperref} 

\usepackage{comment}

\setstcolor{red}

\begin{document}

\title{Time-resolved spectral diffusion of a multimode mechanical memory}\thanks{This work was published in \href{https://doi.org/10.1103/tcpy-jhjy}{Phys.\ Rev.\ Lett.\ \textbf{134}, 243604 (2025).}}

\author{Niccol\`{o} Fiaschi}
\email{nfiaschi@berkeley.edu}
\affiliation{Kavli Institute of Nanoscience, Department of Quantum Nanoscience, Delft University of Technology, 2628CJ Delft, The Netherlands}
\author{Lorenzo Scarpelli}
\affiliation{Kavli Institute of Nanoscience, Department of Quantum Nanoscience, Delft University of Technology, 2628CJ Delft, The Netherlands}
\author{Alexander Rolf Korsch}
\affiliation{Department of Physics, Fudan University, Shanghai 200433, P.R. China}
\affiliation{Department of Physics, School of Science, Westlake University, Hangzhou 310030, P.R. China}
\affiliation{Kavli Institute of Nanoscience, Department of Quantum Nanoscience, Delft University of Technology, 2628CJ Delft, The Netherlands}
\author{Amirparsa Zivari}
\affiliation{Kavli Institute of Nanoscience, Department of Quantum Nanoscience, Delft University of Technology, 2628CJ Delft, The Netherlands}
\author{Simon Gr\"oblacher}
\email{s.groeblacher@tudelft.nl}
\affiliation{Kavli Institute of Nanoscience, Department of Quantum Nanoscience, Delft University of Technology, 2628CJ Delft, The Netherlands}


\begin{abstract}
	High-frequency phonons hold great promise as carriers of quantum information on-chip and as quantum memories. Due to their coherent interaction with several systems, their compact mode volume, and slow group velocity, multiple experiments have recently demonstrated coherent transport of information on-chip using phonon modes, interconnecting distinct quantum devices. Strongly confined phonons in waveguide-like geometries are particularly interesting because of their long lifetime. However, spectral diffusion has been observed to substantially limit their coherence times~\cite{Meenehan2014,Wallucks2020,MacCabe2020}. Coupling to two-level systems is suspected to be a major contributor to the diffusion; however, to date, the origin and underlying mechanisms are still not fully understood. Here, we perform a time-domain study on two adjacent mechanical modes (separated by around $\SI{5}{MHz}$) and show that the frequency positions of the two modes are not correlated in time, in agreement with our theoretical model and Monte-Carlo simulations. This result is an important step in fully understanding the microscopic mechanisms of dephasing in mechanical quantum buses and memories.	
\end{abstract}

\maketitle

\section*{Introduction}

Over the past years, several groundbreaking experiments have shown that high-frequency phonons can be effectively used as carriers of quantum information on-chip. In particular, surface acoustic waves (SAWs) have enabled quantum state transfer and remote entanglement between two superconducting qubits~\cite{Bienfait2019}. SAWs have received most of the early interest since they represent a mature platform for coherent control of quantum information on-chip, having also allowed to realize a beam-splitter~\cite{Qiao2023} and a phase shifter~\cite{Shao2022}. However, SAWs have inherently limited coherence time due to their short lifetime on the order of $\mu$s~\cite{Andersson2021}. More strongly confined phonons in waveguide-like systems are a promising alternative to overcoming these limitations and allow for on-chip distribution of quantum information~\cite{Zivari2022,Zivari2022s}. In this type of system, the coherence of the information that travels in the waveguide is ultimately limited by the spectral diffusion, i.e.\ the frequency jitter of the mechanical mode. This phenomenon was initially observed and studied in single-mode optomechanical cavities~\cite{Meenehan2014}, where the coherence time of the single mechanical mode is on the order of $\SI{100}{\micro s}$~\cite{Wallucks2020,MacCabe2020}. The generally accepted interpretation of this phenomenon involves a bath of two-level systems (TLS) that interact with the mechanical mode(s)~\cite{Meenehan2014,MacCabe2020}. Similar types of defects are widely studied in superconducting circuits since they are a major source of energy loss and decoherence~\cite{Klimov2018,mittal2023,Bozkurt2023, zhang2024}. Engineering the phononic environment that interacts with the TLS can be used to investigate and control the TLS~\cite{chen2024, Odeh2025}. While their actual nature remains elusive, efforts in understanding the microscopic characteristics of these TLS are ongoing~\cite{Mueller2019}.

In this work, we take a new approach to studying the time dynamics of the spectral diffusion of two spectrally close mechanical modes (centered at $\sim$$\SI{5}{GHz}$ and spaced by $\sim$$\SI{5}{MHz}$) of the same optomechanical device, with a time resolution of a few $\mu s$. We find that the frequency positions of the two mechanical modes are not correlated in time. Even with the measurements operating in a continuous fashion, leading to an effective thermal bath elevated from the \SI{50}{mK} environment, our result helps in understanding the effect of TLSs on mechanical resonators, developing a method for fast time resolution of TLSs induced frequency shifts. To support our findings and for a better understanding of the TLS to mechanical modes interaction, we further develop a theoretical model and perform Monte-Carlo simulations of the frequency shift -- in good qualitative agreement with the measured data. The model confirms the coupling mechanism, strength and TLSs spectral distribution. The developed method can help the design of mechanical systems, providing a tool to estimate the spectral diffusion for any mechanical mode.

\section*{Methods}

\begin{figure*}[ht!]
	\includegraphics[width = 1\linewidth]{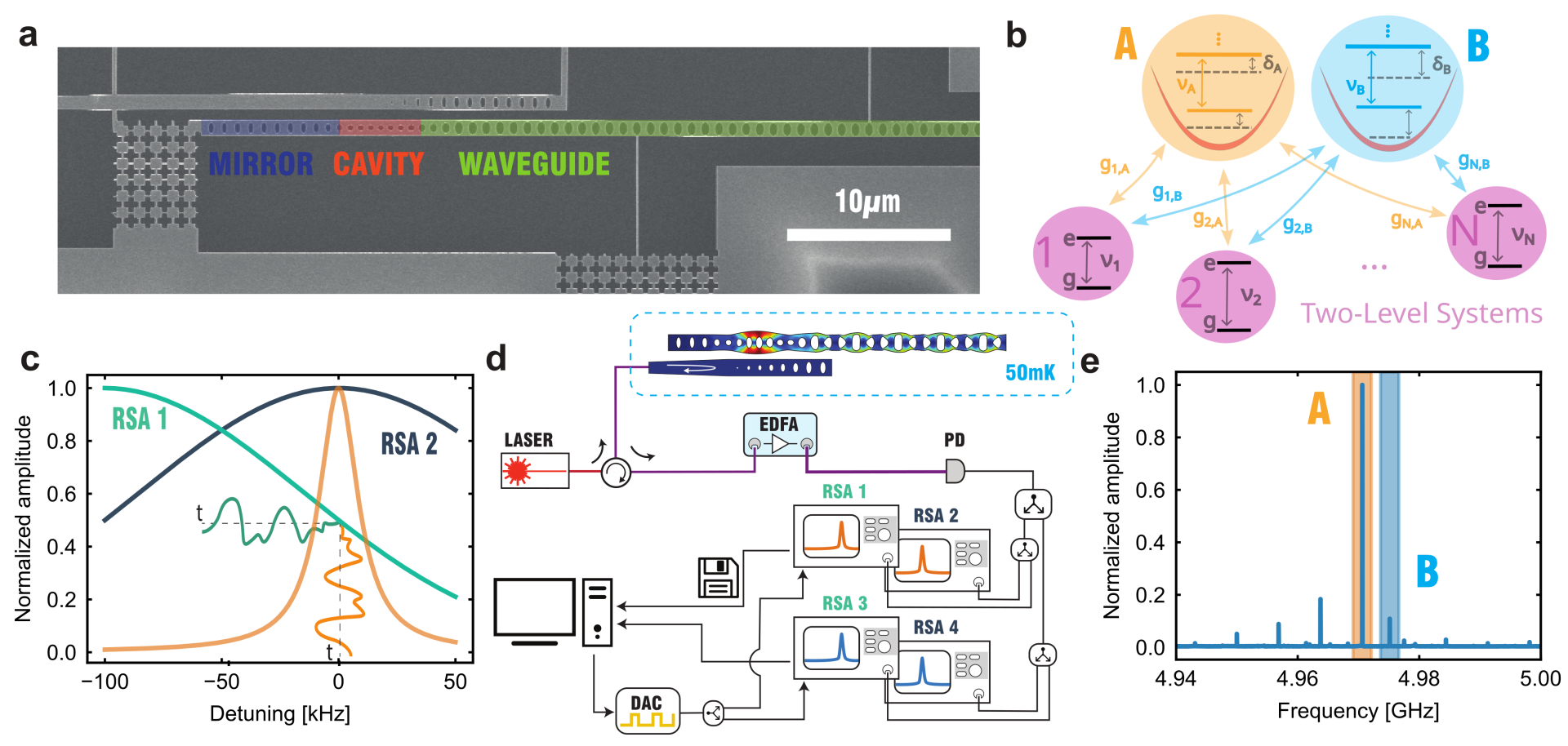}
	\caption{\textbf{Setup and device characterization.} a) SEM image of a device nominally identical to the one used in the measurements. The optomechanical cavity is highlighted in red, the photonic and phononic mirror in blue, and the single-mode (for in-plane symmetric modes) mechanical waveguide (and photonic mirror) in green. The total length of the waveguide $\sim$$\SI{200}{\micro m}$. b) Schematic of the theoretical model for the Monte-Carlo simulations. The mechanical modes A and B at frequency $\nu_{\text{A}}$, $\nu_{\text{B}}$ are coupled to a bath of $N$ TLS. Each TLS has frequency $\nu_{i}$ and coupling rate $g_{i,\text{A}}$, $g_{i,\text{B}}$, with $i$ being the index of the TLS, to mode A and B respectively. Note that the coupling rate for the two modes is, in general, different for each TLS. The bath shifts the frequency of mode A by $\delta_{\text{A}}$ and of B by $\delta_{\text{B}}$. c) Sketch of the method used for resolving the spectral diffusion of a single mode mechanical mode in time (orange line, averaged PSD). See the text for details. d) Scheme of the experimental setup. The laser is routed to the device at the base plate of the dilution refrigerator via a fiber circulator and a lensed fiber. The light coming out of the device is amplified with an Erbium Doped Fiber Amplifier (EDFA) and sent to a fast photodiode. The signal from the photodiode is split with RF splitters and analyzed with four synchronized Real-Time Spectrum Analyzers (RSA1, 2, 3, and 4). All RSAs are triggered externally via a TTL signal from a DAC card, which allows sub-$\mu s$ synchronization. e) Mechanical spectrum of the device with several visible modes. The modes measured in this paper are called mode A (shaded in orange) and mode B (shaded in blue).}
	\label{Fig:1_jitter}
\end{figure*}

The device used in this work, identical to the one of~\cite{Zivari2022}, consists of an optomechanical cavity coupled to a phononic waveguide, where the free-standing end acts as a mechanical mirror. The hybridized mechanical modes of the structure form a series of Fabry-P\'{e}rot modes, with similar spatial profile, closely spaced by $\sim$$\SI{5}{MHz}$ around a central frequency of $\sim$$\SI{5}{GHz}$. The optical cavity features a single mode for the optical field with a resonance in the telecom band around $\sim$$\SI{1550}{nm}$. Fig.~\ref{Fig:1_jitter}a shows a scanning electron microscope (SEM) image of a device nominally identical to the one used in this work.  We model our system as two mechanical modes of the cavity-waveguide system (modes A and B), which are dispersively coupled to a bath of TLS. The dispersive coupling results in a frequency shift of the mechanical modes, which depends on the instantaneous occupation, coupling strength, and detuning of the TLS to that particular mode. Differences in coupling strengths and detunings for the two modes result in different frequency shifts and lead to uncorrelated frequency jitters (see Fig.~\ref{Fig:1_jitter}b for a schematic of the model and the SI section 9~\cite{SI_nfiaschi2025} for more details).

\begin{figure*}[ht!]
	\centering
	\includegraphics[width = 1\linewidth]{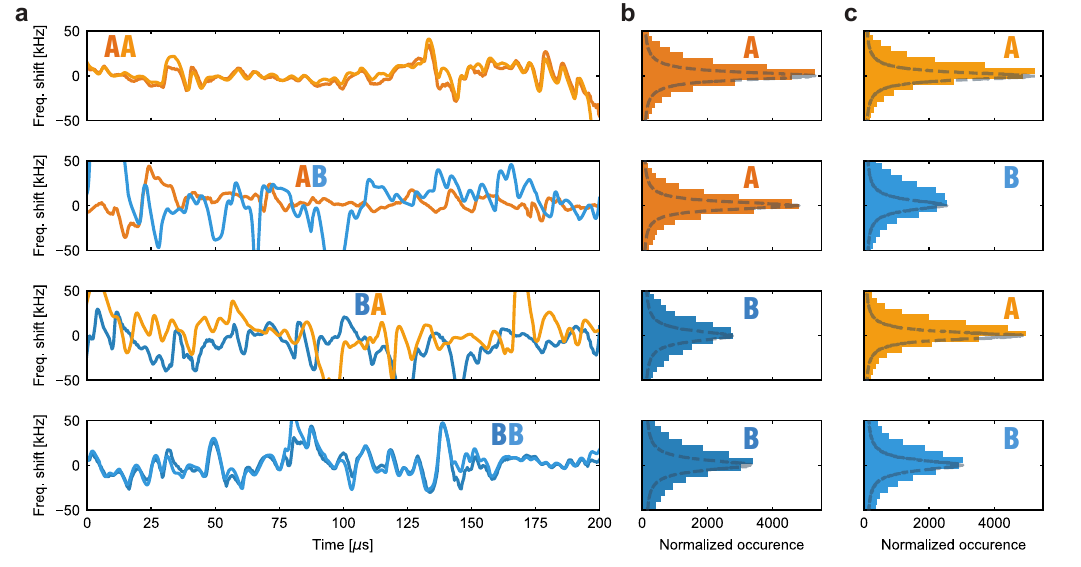}
	\caption{\textbf{Frequency jitter in time.} a) Time traces for mode A (orange measured on RSA1 and light orange on RSA3) and B (blue measured on RSA1 and light blue on RSA3). From top to bottom:\ the synchronized measurements are for the configurations AA, AB, BA, and BB measured from RSA1 and RSA3, respectively (using RSA2 and RSA4 for the amplitude normalization). AA and BB have clearly (almost) identical time traces, while AB and BA show uncorrelated time traces. b) and c) Histogram of the frequency shift modes AA, AB, BA, BB (from top to bottom) for a trace with \SI{10}{ms} length. The dashed black line is the averaged mechanical spectrum (independently) measured with the corresponding RSA, which is in good agreement with the histogram. All data here are measured at \SI{50}{mK}.}
	\label{Fig:2_jitter}
\end{figure*}

To resolve the mechanical frequency of the device in time, we use a frequency demodulation scheme, with its main working principle illustrated in Fig.~\ref{Fig:1_jitter}c. For each mode, we first measure its power spectral density (PSD) and average it over several scans (orange curve). We then use two real-time spectrum analyzers (RSAs) in zero-span mode to acquire the total power filtered from the resolution bandwidth (RBW) filter as a function of time. One RSA (RSA1, with RBW depicted by the green line in the figure), has the filter detuned by -RBW/2 = -\SI{100}{kHz} from the mechanical mode. In this way, the frequency jitter is mapped onto an amplitude modulation of the signal measured by RSA1. The dark orange line depicts a sketch of the time trace of the frequency jitter, while the dark green line is the corresponding amplitude variation of the signal measured (and the dashed lines are the time axes). As the measured power is also proportional to the thermal mechanical population, a change in the amplitude of the mechanical peak will also cause a change in the amplitude of the measured signal. For this reason, the second RSA (RSA2, RBW in dark gray) has the filter on resonance with the mechanical mode, such that its sensitivity to the frequency jitter is negligible. In this way, we can measure the amplitude fluctuations of the signal with RSA2 and the amplitude and frequency fluctuations with RSA1. Normalizing the signal from RSA1 with the one from RSA2 then allows us to extract the frequency position of the mechanical mode in time, even if the signal is modulated in amplitude and frequency. We use two RSAs in this configuration for each mechanical mode in all the measurements.

The setup for the frequency demodulation is schematically shown in Fig.~\ref{Fig:1_jitter}d. A continuous wave (CW) laser, red detuned from the optical resonance by the average mechanical frequency of the modes of interest ($\approx\SI{5}{GHz}$), is routed to the device at the mixing chamber of a dilution refrigerator (temperature of around \SI{50}{mK}) via a circulator and a lensed fiber (coupling efficiency of 10\%). The CW light is partially absorbed by the device, creating a thermal population in the mechanical modes of interest. This population is read via the beam-splitter interaction with the same CW laser, which creates sidebands detuned from the laser carrier by the mechanical frequencies~\cite{Zivari2022}. Throughout all measurements, the instantaneous photon number in the optomechanical cavity is approx.\ 3,000 (see SI section 1~\cite{SI_nfiaschi2025} for more information). The output of the circulator is amplified with an erbium doped fiber amplifier (EDFA) to boost the signal on a fast photodetector. The output of the detector is split with three power splitters and sent to 4 RSAs (1-4). RSA1 and 3 are detuned by -RBW/2 = -\SI{100}{kHz} from the measured mechanical mode, while RSA2 and 4 are always on resonance with them. The RSAs are synchronized in the data acquisition via a TTL signal that triggers all the RSAs simultaneously, with a relative delay of sub-$\mu$s.

After the scan of the optical resonance (cf.\ SI section 1~\cite{SI_nfiaschi2025}), we measure the PSD of the device over a broad range of frequencies to identify the two mechanical modes of interest, which is plotted in Fig.~\ref{Fig:1_jitter}e. We see that the device has a comb of mechanical modes due to the hybridization of the single-mode mechanical cavity with the Fabry-P\'{e}rot cavity formed by the long waveguide with its free-standing end~\cite{Zivari2022}. The shaded regions in the plot highlight the two modes used in this work. The resonances have a FWHM = \SI{20}{kHz} (see SI section 1~\cite{SI_nfiaschi2025} for details). We use optomechanical induced transparency (OMIT) to determine the optomechanical coupling of each mode, finding \SI{0.3}{MHz} for A and \SI{0.2}{MHz} for B. We also measure a phonon lifetime of around $\SI{1}{ms}$ (see SI section 2~\cite{SI_nfiaschi2025}).

For an initial characterization of the frequency diffusion, we use RSA1 and RSA3 to take synchronized fast frequency scans of the peaks (PSD), similar to the procedure reported in~\cite{Wallucks2020}. To obtain sufficient information of the peak, the fastest possible scan takes $\SI{200}{\micro s}$ (RBW = \SI{10}{kHz}). In this configuration, approximately half of the scans have more than one prominent peak (see SI section 3~\cite{SI_nfiaschi2025}). This indicates that the time scale of the jitter is much faster than the scanning time. We overcome this limitation by using a frequency demodulation scheme, which allows to use larger RBW and, therefore, have finer time resolution.

\section*{Results}

We use the four RSAs to measure (synchronized) time traces of the two mechanical modes in four configurations:\ AA, AB, BA, and BB. RSA1 (3) is detuned by $-\textrm{RBW}/2 = -\SI{100}{kHz}$ from the mode A (A), A (B), B (A) and B (B), respectively, while RSA2 and 4 are always on resonance with the mode measured by RSA1 and 3, respectively. Each time trace is \SI{10}{ms} long. For this measurement the choice of the RBW is crucial:\ we set a \SI{200}{kHz} RBW filter as a trade-off for having a large enough time resolution ($\sim$$1/\textrm{RBW} = \SI{5}{\micro s}$) while still maintaining a good signal to noise ratio, high dynamical range and high sensitivity. The RBW is also chosen to be much bigger than the average frequency shift to avoid signal distortion. We report an average signal-to-noise ratio of 10 for mode A (with a maximum value of 80) and 5 for B (with a maximum value of 30). The peak fluctuates in the linear part (in the detuned case) and on the flat part (in the resonant case) of the filter. For the chosen filter, a typical frequency fluctuation of \SI{10}{kHz} (FWHM/2) gives a 0.7\% relative change in amplitude for the filter at resonance with the mechanical mode and a 7\% relative change for the filter detuned by $-\textrm{RBW}/2$. With the amplitude of the signal from RSA2 and 4, we normalize the trace measured from RSA1 and 3 (see SI section 6~\cite{SI_nfiaschi2025}). We then use the filter shape to convert the amplitude fluctuations into frequency fluctuations (see SI section 5~\cite{SI_nfiaschi2025}). The result of the measurements is reported in Fig.~\ref{Fig:2_jitter}a, where we plot the frequency shift in time, from top to bottom, for the four configurations AA, AB, BA, and BB, in orange (blue) and light orange (light blue) for mode A (B). In this plot, we show a $\SI{200}{\micro s}$ long segment of the measured trace. It is clearly visible that the AA and BB configurations have highly correlated traces, while the AB and BA configurations have uncorrelated time traces of the two modes. The (common mode) frequency jitter, which can be caused by laser frequency shifts and changes in laser power, is much smaller than the measured frequency shifts (see SI section 4~\cite{SI_nfiaschi2025}). We further show in Fig.~\ref{Fig:2_jitter}b and c, the histogram of the frequency shift for the complete time traces. The letters in the plot refer to which mode is analyzed. The dashed black line is the mechanical PSD as measured from the RSAs in scanning mode (averaging for approx.\ $\SI{30}{s}$ for each measurement). Note that the two (independent) measurements are in very good agreement, confirming the validity of the method used. The small discrepancy in the FHWM of the distribution should arise mainly from the noise in the measurement of the amplitude from RSA2 and 4. The small discrepancy in the center of the distribution is attributed to slow frequency drifts of the mechanical modes~\cite{Wallucks2020}.

\begin{figure}[]
	\centering
	\includegraphics[width = 1\linewidth]{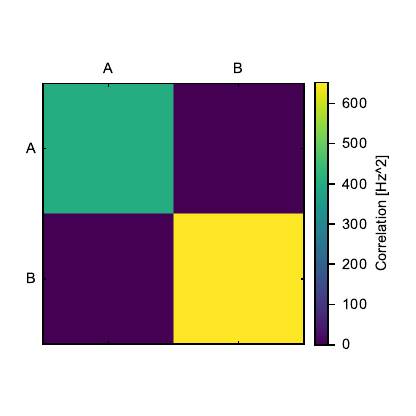}
	\caption{\textbf{Correlation of the frequency jitter.} Correlation for zero delay ($\tau = 0$, as defined in SI section 7~\cite{SI_nfiaschi2025} and~\cite{Kubo1969}) for the mechanical mode configurations AA, AB, BA, and BB using the full traces of Fig.~\ref{Fig:2_jitter} with \SI{10}{ms} length. The AA and BB combinations show strong correlations, while the AB and BA combinations are uncorrelated. All data here are measured at \SI{50}{mK}.}
	\label{Fig:3_jitter}
\end{figure}

To quantitatively assess if the time traces are correlated or not, we calculate the correlation function in the four configurations (see SI section 7~\cite{SI_nfiaschi2025} and~\cite{Kubo1969}), and we show the results for zero delay ($\tau = 0$) in Fig.~\ref{Fig:3_jitter}. It is clearly visible from the matrix that AA and BB have a strong correlation, while AB and BA have uncorrelated values. We would like to note that a more detailed study with different RBWs could be beneficial in extracting short-delay coherence time. To validate the measurements and gain some additional information on the frequency jitter mechanism, we develop a theoretical model and use it to run Monte-Carlo simulations to obtain the frequency diffusion of two mechanical modes shifted by a bath of TLS (see SI section 9~\cite{SI_nfiaschi2025}). For our model, we use a COMSOL simulation to calculate the coupling rates between the mechanical modes of the device and each TLS of the bath. We find that each mechanical mode can couple very differently to the same TLS bath. We then calculate, for each time step of the simulation, the frequency shift of the mechanical mode arising from the dispersive coupling with each TLS. We find good qualitative agreement between the measured data and the model, finding a frequency jitter in time with amplitude and time-scale comparable with the one measured and with the same pattern of correlation, as can be seen in Fig.~\ref{Fig:2_jitter}a and SI section 9~\cite{SI_nfiaschi2025} Fig. 8d, and Fig.~\ref{Fig:3_jitter} and SI section 9~\cite{SI_nfiaschi2025} Fig. 9. 

In order to get a better understanding of the nature of the TLS, we perform the same experiment at a temperature of \SI{800}{mK} (see SI section 8~\cite{SI_nfiaschi2025}), and perform Monte-Carlo simulations up to \SI{4}{K} (see SI section 9~\cite{SI_nfiaschi2025}), and in both cases the correlation results are practically identical. This indicates that a considerable part of the frequency diffusion is caused by far detuned TLS, which are only partially affected by the higher temperature~\cite{MacCabe2020}, as the TLS with frequency close to the mechanical modes of interest are mostly saturated.

\section*{Discussion}

In our experiment, we show that two mechanical modes, even if spectrally close and with similar mode shapes, have an uncorrelated spectral diffusion for time scales greater than a few microseconds. This leads us to conclude that when using two devices in a dual-rail scheme~\cite{Riedinger2018,Fiaschi2021} or a single device with multiple mechanical modes~\cite{Zivari2022s} the resulting coherence times would be similar. The uncorrelated traces and the developed model show that the two modes are coupled very differently to the same surface defect (TLS) bath, which, most likely, are the cause of the jittering~\cite{MacCabe2020}. From the model, we estimate that each TLS can have a difference in coupling rate to the two mechanical modes of up to two orders of magnitude. Our results shed further light on the cause of the frequency diffusion in mechanical oscillators and could be used to design higher coherence time multi mode devices (that have mostly common mode frequency shift, for example). Additional experiments are, however, required to fully understand the underlying mechanisms and to open the way to long coherence-time mechanical quantum memories without the use of dynamical decoupling methods. In particular, a detailed and systematic analysis via surface treatments of the silicon is necessary, or the use of alternative materials such as GaP or silicon nitride. We can further conclude that, in order to reach long coherence times in mechanical quantum memories, it is necessary to saturate the TLS over a very wide frequency range ($\gtrsim\SI{20}{GHz}$), for example, through an AC electric field. For this reason, the use of acoustic metamaterials --phononic shields--, given their limited bandgap ($\sim \SI{2}{GHz}$), have been only partially beneficial for the coherence time~\cite{MacCabe2020, Wallucks2020}.

\medskip

\textbf{Acknowledgments}
We would like to thank Robert Stockill and Raymond Schouten for valuable discussions and Clinton Potts for support. We further acknowledge assistance from the Kavli Nanolab Delft. This work is financially supported by the European Research Council (ERC CoG Q-ECHOS, 101001005) and is part of the research program of the Netherlands Organization for Scientific Research (NWO), supported by the NWO Frontiers of Nanoscience program, as well as through a Vrij Programma (680-92-18-04) grant.

\textbf{Conflict of interests:}\ The authors declare no competing interests.

\textbf{Author contributions:}\ N.F.\ devised and planned the experiment, built the setup and performed the measurements, with help from A.R.K. A.Z.\ fabricated the device. N.F., L.S.\ and S.G.\ analyzed the data and wrote the manuscript. S.G.\ supervised the project.

\textbf{Data Availability:}\ Source data for the plots are available on \href{https://doi.org/10.5281/zenodo.15584323}{Zenodo}~\cite{Zenodo_nfiaschi2025}.

\bibliographystyle{apsrev4-2}

\setcounter{figure}{0}
\renewcommand{\thefigure}{S\arabic{figure}}
\setcounter{equation}{0}
\renewcommand{\theequation}{S\arabic{equation}}

\clearpage

\section*{Supplementary Information}
\label{SI}

\subsubsection{Optical and mechanical resonances}
\label{Optical resonance}

\begin{figure*}[ht!]
	\centering
	\includegraphics[width = 1\linewidth]{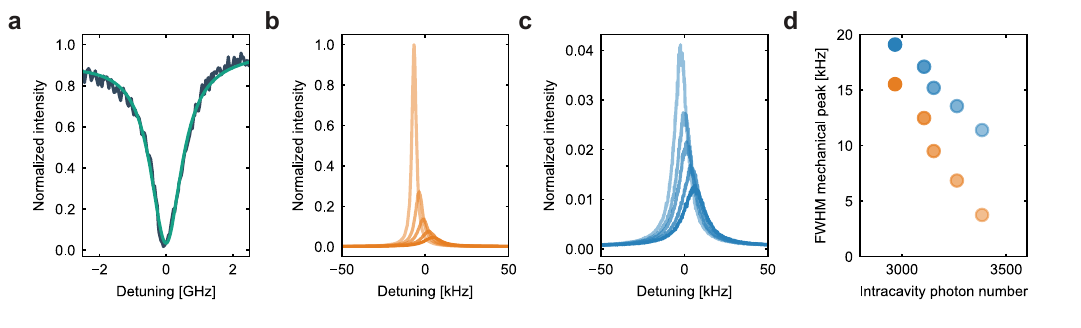}
	\caption{a) Optical resonance of the optomechanical cavity. The resonance is centered at roughly $\SI{1535}{nm}$ and has a linewidth of $\SI{1.1}{GHz}$. b) and c) Averaged mechanical PSD of mode A (in orange) and B (in blue), where each measurement is integrated for $\SI{30}{s}$. Each spectrum is measured with different intracavity photon number (increasing intracavity photon number for lighter colors, see d for the actual values). Optomechanical spring effect and phonon lasing, shift the resonances and decrease the linewidth with increased intracavity photon number. d) Linewidth of the mechanical peaks (A in orange and B in blue), as a function of the intracavity photon number. In the measurement in the main text the intracavity photon number is about $3,000$.}
	\label{Fig:S8_jitter}
\end{figure*}

To measure the optical resonance we scan the laser frequency and measure the reflected power from the device on a photodiode (see Fig.~\ref{Fig:S8_jitter}a). From a Lorentzian fit we can extract a resonance wavelength of around $\SI{1535}{nm}$ and a linewidth of $\SI{1.1}{GHz}$. We then measure the (averaged) PSD of the two mechanical modes, Fig.~\ref{Fig:S8_jitter}b and c, as a function of the intracavity photon number, which is proportional to the power sent to the device (see Fig.~\ref{Fig:S8_jitter}d for the actual values). As expected, the optomechanical spring effect and phonon lasing shift the mechanical resonances and decrease the linewidth with increased intracavity photon number~\cite{Aspelmeyer2014,Meenehan2014,Cui21}. The measured mechanical linewidths as a function of the intracavity photon number are shown in Fig.~\ref{Fig:S8_jitter}d. In the measurement in the main text the intracavity photon number is around 3,000, which is the lowest value we could use to still obtain a high enough signal to noise ratio for the time traces of mode B (the mode with lower optomechanical coupling, $g_0$).

\subsubsection{Lifetime}
\label{lifetime}

\begin{figure}[h!]
	\centering
	\includegraphics[width = 1\linewidth]{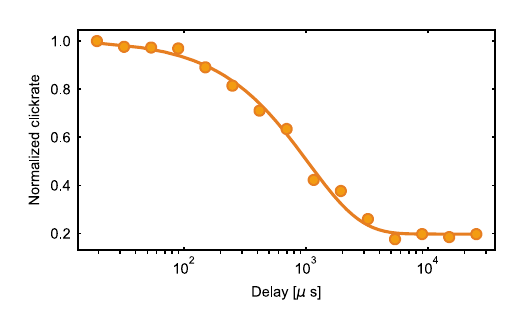}
	\caption{Normalized click rates of the probe pulse as a function of the delay between the two pulses (pump and probe). The rates are directly proportional to the thermal population of the device allowing to measure the lifetime of the phonons in the structure. From the exponential fit (orange line) we extract a $T_1 \approx \SI{1}{ms}$. }
	\label{Fig:S1_jitter}
\end{figure}

To determine the lifetime of the phonons in the device we send a series of pairs of red-detuned pulses (pump and probe~\cite{Fiaschi2021}). The pump pulse (the first one to arrive at the device), creates a thermal population in the cavity, while the probe pulse measures the thermal occupation of the cavity. We repeat the pair of pulses every \SI{50}{ms}. Sweeping the delay between the pulses, we measure the decay of the thermal population created from the first pulse and extract the lifetime of the phonons. The data are shown in Fig.~\ref{Fig:S1_jitter}, with a measured lifetime of $T_1 \approx \SI{1}{ms}$.

\subsubsection{Fast RSA scans}
\label{Fast scans on the RSA}

\begin{figure*}[ht!]
	\centering
	\includegraphics[width = 1\linewidth]{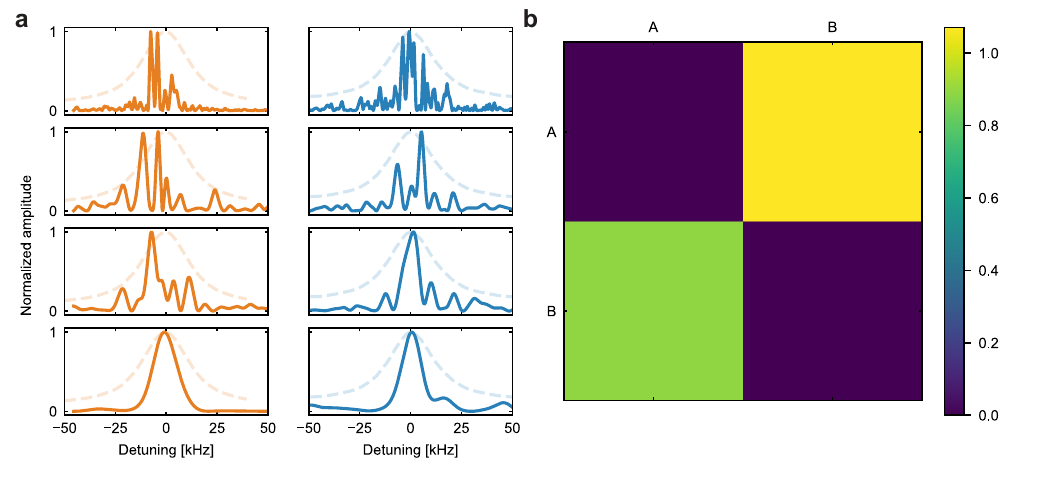}
	\caption{a) Fast scans of the mechanical modes A (in orange) and b (in blue) around their central frequencies using a RBW of \SI{1}{kHz}, \SI{3}{kHz}, \SI{5}{kHz} and \SI{10}{kHz} from top to bottom. The bottom scans are the fastest and have a sweep time of $\approx$$\SI{200}{\micro s}$. It is visible that, most of the time, the scans with finer RBW (and so slower sweep time) measure the mechanical peak several times in a single sweep. For the \SI{10}{kHz} case, most of the time the scans show a single prominent peak. The dashed lines are the mechanical spectrum with 3000 averages. b) Correlation distance matrix between the modes A and B. The AA and BB correlations are very close to zero, showing a very strong autocorrelation, while the AB and BA are very close to unity, showing that are uncorrelated. }
	\label{Fig:S2_jitter}
\end{figure*}

The correlations of the frequency positions of the two mechanical modes can be measured by performing synchronized fast scans on the RSAs. Ideally, if the jitter is slower than the sweep time of the scan, only one peak should be visible. From this measurement the positions of the peak can be measured and the correlation can be calculated. As shown in Fig.~\ref{Fig:S2_jitter}a, we see that for small RBW the spectra mostly exhibit multiple peaks however, indicating that the scan sweeps are too slow. For large  RBW of around \SI{10}{kHz} our scan is fast enough (with a sweep time of $\SI{200}{\micro s}$) and has sufficient resolution to distinguish single peaks inside the average envelope. With this last bandwidth we take 300 scans for each combination of peaks (AA, AB, BA and BB) and post-select the ones with a clear single peak, obtaining between 100 and 150 single peaks for each combination. We use the maximum values of the peaks for each pair of scans as measurement of the mechanical frequencies, creating two vectors. We then use the correlation distance between the vectors to asses if the mechanical frequencies are correlated. For two vectors $\vec{u}$ and $\vec{v}$ we use the correlation distance $C$ as defined in the Python package SciPy

\begin{equation*}
	C = 1-\frac{(\vec{u}-\vec{\bar{u}}) \cdot (\vec{v}-\vec{\bar{v}})}{|| \vec{u}-\vec{\bar{u}} ||_2  || \vec{v}-\vec{\bar{v}} ||_2}.
\end{equation*}

With this definition, two perfectly correlated (anti-correlated) vectors have a correlation distance of 0 (2), while two uncorrelated vectors have a correlation distance of 1. We report the results of the correlations between the measured peak positions in Fig.~\ref{Fig:S2_jitter}b, where we show that the combination AA and BB are correlated while AB and BA are uncorrelated.

\subsubsection{Common mode jitter}
\label{Common mode jitter}

\begin{figure*}[ht!]
	\centering
	\includegraphics[width = 1\linewidth]{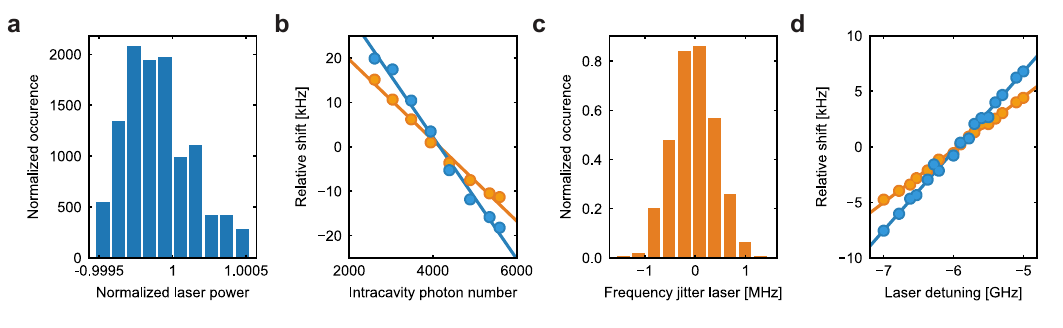}
	\caption{a) Histogram of the normalized power of the laser. The standard deviation of the distribution is approx.\ $2\times10^{-4}$. The measurement was taken with an integration time of \SI{100}{ms} and lasted in total one minute, much longer than the time of a single measurement reported in the main text. b) Frequency shift of the mechanical peaks as a function of the intracavity photon number, for mode A (orange) and mode B (blue). From the linear fit (solid line, same color as the datapoints) and the distribution of laser power reported in a), we can extract a maximum (common mode) frequency jitter given by the power fluctuation of the laser of $\SI{20}{Hz}$, much smaller than the frequency jitter of the mechanical modes. c) Normalized histogram of the frequency jitter of the laser. The standard deviation of the distribution is approx.\ $\SI{0.42}{MHz}$. The measurement was taken with a rate of 5~Hz and lasted 3 minutes, again much longer than the time of the single measurements reported in the main text, which leads to an upper bound of the laser frequency shift. d) Frequency shift of the mechanical frequencies of mode A (orange) and B (blue) as a function of the laser detuning from the optical resonance. From the linear fit (solid line, same color as the datapoints) and the distribution of laser frequencies reported in c), we can extract a maximum (common mode) frequency jitter given by the frequency fluctuations of $\SI{30}{Hz}$, also much smaller than the frequency jitter of the mechanical modes. In both b) and d) the differences in the slopes are given by the different $g_{0}$ of the two modes.}
	\label{Fig:S3_jitter}
\end{figure*}

Fluctuations in laser power and detuning from the optical resonance cause a (common mode) frequency shift of the mechanical modes. We estimate this jitter using the distribution of laser power and frequencies. We measure the laser power (renormalized to 1) and its frequency in time and we report the normalized histogram of these two quantities in Fig.~\ref{Fig:S3_jitter}a and c. The standard deviations of the distributions are approx.\ $2\times10^{-4}$ and $\SI{0.42}{MHz}$, respectively. We also measure the mechanical frequencies as function of the intracavity photon number and as function of the detuning of the laser from the optical resonance. As shown in Fig.~\ref{Fig:S3_jitter}b and d, we can extract a maximum (common mode) frequency jitter of the mechanical modes of $\approx\SI{20}{Hz}$ given by the laser power fluctuations, and of $\approx\SI{30}{Hz}$ given by the laser frequency fluctuations. Theses frequency shifts are much smaller than the one reported in the main text and can therefore not explain the (uncorrelated) frequency diffusion of the modes.

\subsubsection{Characterization of RSA filter}
\label{Calibration of RSA filter}

\begin{figure}[h!]
	\centering
	\includegraphics[width = 1\linewidth]{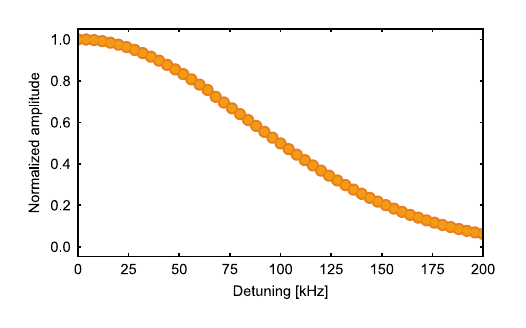}
	\caption{Average value of \SI{1}{ms} traces with different detuning of the filter center with respect to a microwave source, for a RBW of \SI{200}{kHz}. From the Gaussian fit, we obtain a FWHM of \SI{200}{kHz} for all RSAs.}
	\label{Fig:S4_jitter}
\end{figure}

It is critical to know the filter shape of the RSA precisely, in order to being able to convert faithfully from the amplitude measurement to a frequency. We use an external microwave source at a (fixed) frequency around \SI{5}{GHz}. We measure time traces of \SI{1}{ms} at several detunings of the filter from the microwave frequency and then average the traces and plot them as a function of the detuning in Fig.~\ref{Fig:S4_jitter}. From the Gaussian fit we find a FWHM of the filter functions of about \SI{200}{kHz} for all RSAs used, consistent with the RBW of \SI{200}{kHz}.

\subsubsection{Amplitude fluctuations}
\label{Amplitude fluctuations}

\begin{figure*}[ht!]
	\centering
	\includegraphics[width = 1\linewidth]{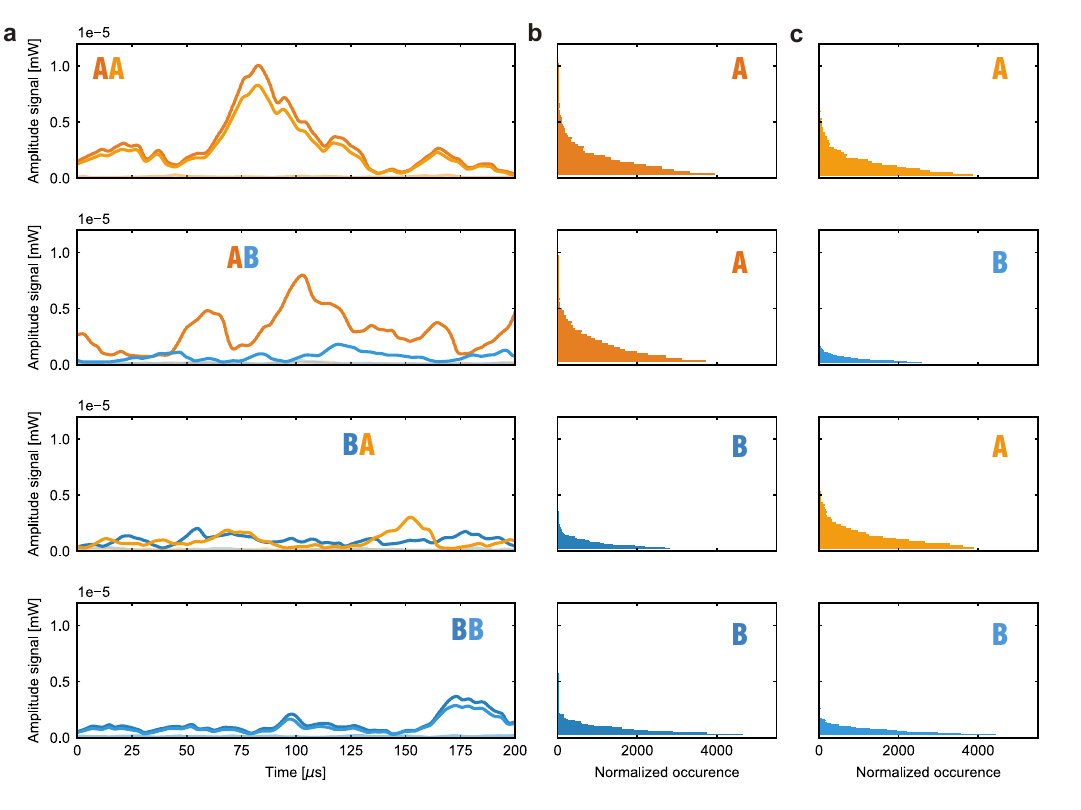}
	\caption{a) Time traces of the amplitude fluctuations of the signal for mode A (orange, RSA2 and light orange, RSA4) and B (blue, RSA2 and light blue, RSA4) with the filters on resonance with the mechanical modes. From top to bottom:\ the synchronized measurements are for the configurations AA, AB, BA and BB measured with RSA2 and RSA4 respectively, and are for the same time as the data in Fig.~\ref{Fig:2_jitter}. AA and BB have (almost) identical time traces, while AB and BA show uncorrelated time traces even for the amplitude of the modes. These fluctuations are the time-resolved Brownian thermal motion. The small difference in signal comes from different losses and splitting ratios of the RF power splitters and is compensated in post-analysis to convert the y-axis to frequency in Fig.~\ref{Fig:2_jitter}. The partially transparent traces are the background measurements taken \SI{200}{MHz} from the mechanical modes of interest (at a frequency with no visible peaks). b) and c) Histogram of the amplitude fluctuations as measured from RSA2 and RSA4 for the modes AA, AB, BA, BB (from top to bottom) for a trace with \SI{10}{ms} length. The histograms follow the probability distribution of a thermal state.}
	\label{Fig:S5_jitter}
\end{figure*}

In our measurements we consistently see that the amplitude of the signals fluctuates in time, even in the case of the filter of the RSA on resonance with the mechanical modes. In Fig.~\ref{Fig:S5_jitter}a we report a time trace (for the same time as Fig.~\ref{Fig:2_jitter}), for the four configurations (AA, AB, BA and BB) as measured from RSA2 and RSA4 with the filter on resonance. In this configuration, from the frequency jitter, we can expect a 0.7\% change in amplitude for a frequency fluctuation of FWHM/2. However, in the data we see that the amplitude variations span an order of magnitude and are not correlated between the two modes. This clearly excludes common mode amplitude fluctuations which can come from changes in laser power or frequency, amplification of the EDFA, or losses in the optical path (from vibration of the lensed fiber given by the pulse tube of the dilution refrigerator, for example). Since the amplitude of the signal is directly proportional to the (thermal) population of the mode, we are measuring the thermal Brownian motion in time. This is clearly visible as their histograms are an exponential distribution in frequency, perfectly following the probability distribution of a thermal state (cf.\ Fig.~\ref{Fig:S5_jitter}b and c for the full \SI{10}{ms} trace for all four cases).

\subsubsection{Correlation}
\label{correlation SI}

To quantitatively assess the correlations in the four configurations, we calculate correlation functions. For this, we use the complete traces taken from the RSAs, with a total length of \SI{10}{ms}. We then divide each trace into segments, which are defined as subsequent points of the trace that have only physical values of the frequency jitter (nonphysical values can arise from the noisy calibration of the y-axis, particularly when the amplitude measured with RSA2 and RSA4 is comparable with the noise floor). We report that around 2/3 of the total data points have physical values. The correlation function for each segment is defined as
\begin{equation}
	C_j(\tau) = \sum_{i=0}^{N_j}(S_j(t_i)-\hat{S})(S_j(t_i+\tau)-\hat{S}),
\end{equation}
where $S_j$ is the segment $j$, with $N_j$ points, of the frequency shift in time and $\hat{S}$ the average value~\cite{Kubo1969}. We then average between the calculated correlations of all segments (each with a weight given by the number of points of the segment).

\subsubsection{Frequency fluctuations at \SI{800}{mK}}
\label{Frequency fluctuations at 800}

\begin{figure}[ht!]
	\centering
	\includegraphics[width = 1.0\linewidth]{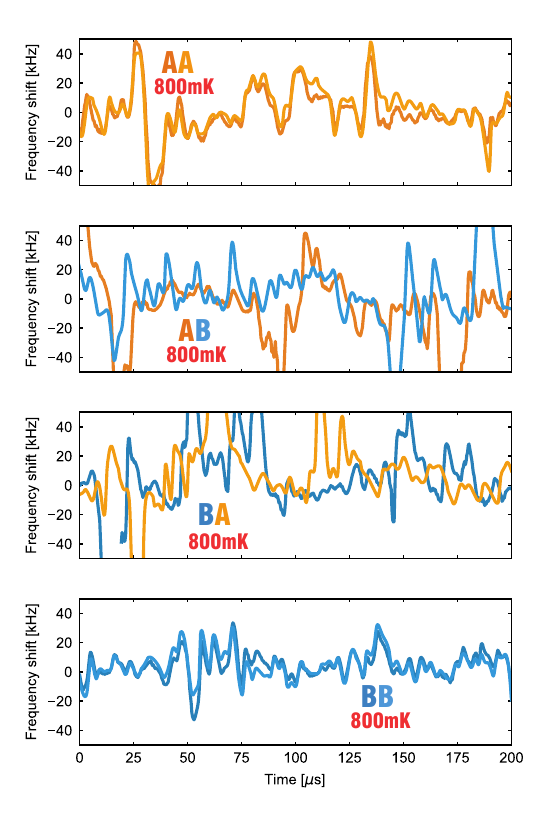}
	\caption{Time traces of the frequency jitter for mode A (orange, RSA1 and light orange, RSA3) and B (blue, RSA1 and light blue, RSA3) with the temperature of the mixing chamber plate of the dilution refrigerator at \SI{800}{mK}. From top to bottom:\ the synchronized measurements are for the configurations AA, AB, BA and BB measured from RSA1 and RSA3 respectively (using RSA2 and RSA4 for the amplitude normalization). The qualitative behavior does not change compared to the measurements at \SI{50}{mK} reported in Fig.~\ref{Fig:2_jitter}.}
	\label{Fig:S6_jitter}
\end{figure}

To study the behavior of the frequency jitter in more detail, we perform the same measurement reported in Fig.~\ref{Fig:2_jitter} at an elevated temperature of the mixing chamber plate of \SI{800}{mK}. As can be seen in Fig.~\ref{Fig:S6_jitter}, the behavior and the correlations are practically identical to the ones measure at \SI{50}{mK}, which we believe is due to the high power of the CW laser used causing heating of the device, and increasing its effective temperature to around \SI{1}{K}.

\subsubsection{Monte-Carlo simulation}
\label{Monte-Carlo}

In order to obtain a better understanding of the underlying TLS dynamics, we perform a Monte-Carlo simulation frequency shifts of two mechanical modes coupled to a bath of TLS. We use COMSOL to calculate the coupling rate of each mechanical mode to the TLS, where each TLS is assumed to dispersively couple to each mechanical mode. In particular, we calculate the TLS state-dependent frequency shifts of each mode for each time step. We then convolute the results with the RSAs filter and calculate the correlation between the time traces. We find that the time traces are only correlated on a very fast time scale, and uncorrelated on a time scale longer than the inverse of the RSAs filter bandwidth of around $\SI{5}{\micro s}$ in our case.

\begin{figure*}[ht!]
	\centering
	\includegraphics[width = 1.\linewidth]{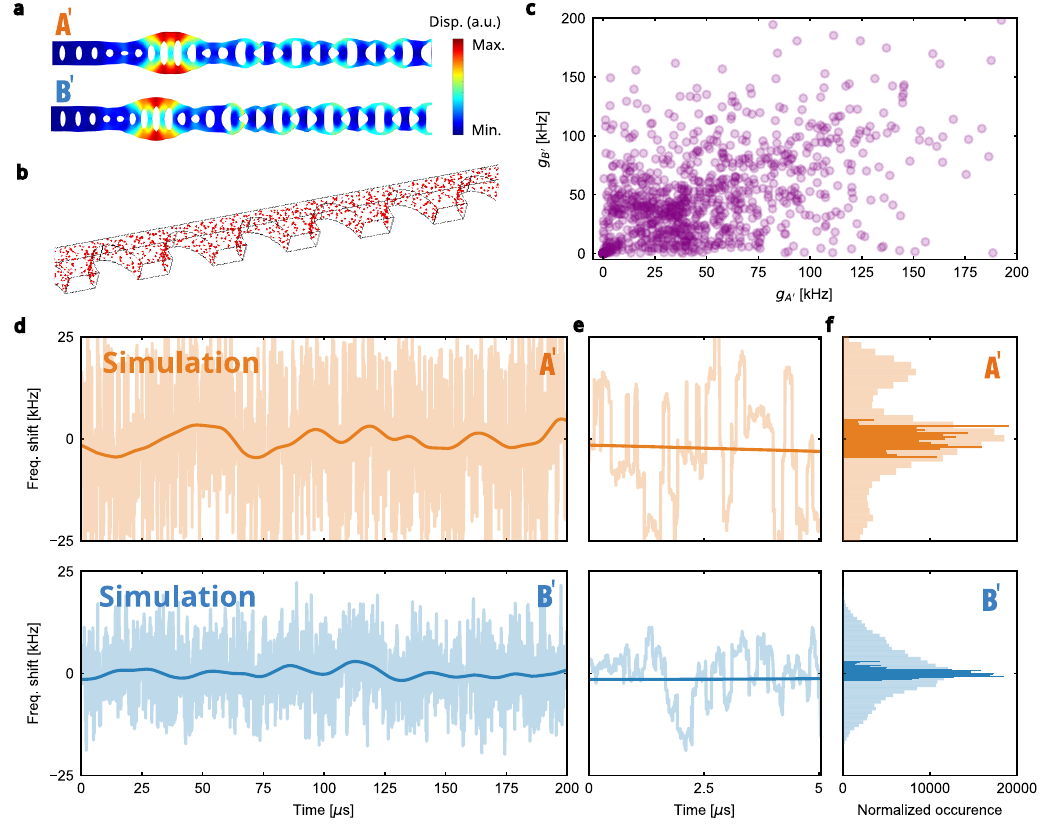}
	\caption{a) Simulations of the displacement of the mechanical modes A' (top) and B' (bottom). Note how the modes have their maximum and minimum displacements in different spatial positions. b) Section of the simulated structure. The red dots are the positions of the points used to evaluate the strain, each corresponding to the positions of a single TLS. They are randomly distributed and have a maximum depth of \SI{10}{nm} from every surface. c) Plot of the calculated coupling strength between mechanical modes A' and B' and the TLS, where each point is a single TLS. For simplicity, we only plot 1,000 points. Note how the coupling strengths $g_{\text{A'}}$ and $g_{\text{B'}}$ can vary substantially. d) Simulated frequency shift of mode A' (in orange, top) and B' (in blue, bottom) due to the dispersive coupling with the TLS bath. The dark lines are the signal with RSA filter (RBW = \SI{200}{kHz}), and the light solid lines are unfiltered. e) Zoom-in of the time traces reported on d, where the fast dynamics of the frequency shift becomes visible. The RSA filter smoothens the telegraphic noise coming from the TLS. f) Histograms of the full simulation (10 trajectories) for mode A' (in orange, top) and B' (in blue, bottom), as before. Again, dark colors are used for the histograms of the time traces passing the RSA filter and light ones for the unfiltered traces. Note that the secondary peaks given by single TLS shifting the frequency of the mechanical modes that are visible in A', are averaged out by the RSA filter.}
	\label{Fig:S7_jitter}
\end{figure*}

We further simulate the mirror, cavity, and waveguide regions of the device using COMSOL. As previously done, the total length of the waveguide is limited to 100 unit cells to reduce the calculation time~\cite{Zivari2022}. The total length of the waveguide part in the simulation is then $\SI{50}{\micro m}$ or about 1/5 of the actual length of the device measured. We simulate two modes (mode A' and B') at frequencies $\nu_{\text{A'}}= \SI{4.847}{GHz}$ and $\nu_{\text{B'}}=\SI{4.870}{GHz}$. These two modes are Fabry-P\'{e}rot modes, just like the ones measured in the main text (cf.\ Fig.~\ref{Fig:S7_jitter}a). Using the density per unit energy and volume of bulk glass ($10^{-45}$ state/J/$\text{m}^3$~\cite{Phillips1987, MacCabe2020}), we calculate a TLS density of around $\SI{600}{states/GHz}$ in the simulated device volume within a depth of \SI{10}{nm} from the surface~\cite{MacCabe2020} (total volume that hosts the TLSs is approx.\ $\SI{1}{\mu m^3}$). We can therefore expect roughly $1.2 \times 10^4$ TLS with frequencies between $\nu_{TLS, min}= \SI{0.5}{GHz}$ and $\nu_{TLS, max} = \SI{20}{GHz}$. Here we limit the upper frequency to \SI{20}{GHz} to reduce the number of interacting TLS and to save on computational time. Due to the slow increase of the maximum frequency shift of the mechanical mode with the chosen maximum TLS frequency, the simulation results are qualitatively similar even for higher $\nu_{TLS, max}$. We randomly choose $1.2 \times 10^4$ (fixed) frequencies $\nu_{i}$ from a uniform distribution between \SI{0.5}{GHz} and \SI{20}{GHz}, where $i$ is the TLS number index. We then randomly choose $1.2 \times 10^4$ points (defined by $x_i$, $y_i$, $z_i$ coordinates) that are (uniformly) distributed on every surface of the simulation with a maximum depth of \SI{10}{nm}. Each point is the position of a single TLS. We show part of the simulated device and the points chosen (red dots) in Fig.~\ref{Fig:S7_jitter}b. Note that, due to the highly symmetric nature of the device, we can simulate only 1/4 of it and use symmetries to get the mechanical mode of the complete device. We evaluate the strain for each point for both mechanical modes A' and B' ($s_{i,\text{A'}}$ and $s_{i,\text{B'}}$). The coupling of each TLS to the modes A' and B' is then defined as $g_{i,\text{A'}} = g_{av}  s_{i,\text{A'}} / s_{av, \text{A'}}$ and $g_{i,\text{B'}} = g_{av}  s_{i,\text{B'}} / s_{av, \text{B'}}$, where $g_{av} =\SI{100}{kHz}$~\cite{Cleland2024, MacCabe2020} and $s_{av,\text{A'}}$, $s_{av,\text{B'}}$ are the average strain for mode A' and B'. We plot the $g_{i,\text{A'}}$ and $g_{i,\text{B'}}$ in Fig.~\ref{Fig:S7_jitter}c for 1,000 of the TLS considered (for simplicity). We note that 40\% of the total TLS have a relative ratio in coupling strength between the two mechanical modes that is larger than 2 (or smaller than 0.5). This is given by the different mode shape, and hence strain, of the two modes, as is visible from Fig.~\ref{Fig:S7_jitter}a. Note that for mechanical modes that have higher spatial overlap integral (i.e., closer frequencies), the difference in coupling strengths to the TLSs will be smaller and the expected cross-correlation will be higher.

To perform the Monte-Carlo simulation, we define the rates at which each TLS decays to the ground state ($\Gamma_{\text{down}}$) and gets excited ($\Gamma_{\text{i,up}}$). We assume all TLS have the same decay rate ($\tau_{\text{down}} = 1/\Gamma_{\text{down}} = \SI{1}{\micro s}$) and an excitation rate given by the thermal distribution ($\Gamma_{i,\text{up}} / \Gamma_{\text{down}} = 1 / (e^{h \nu_{i} / k_B T})$), where T is the bath temperature, $h$ and $k_{B}$ Planck's and Boltzmann's constants, respectively. In this way we have a population of the TLS given by a thermal bath with an average temperature $T$. We choose $\tau_{\text{down}}$ to be close to literature values~\cite{Cleland2024, MacCabe2020}, and smaller than the inverse of the filter bandwidth (1/RBW) since no correlations for longer times have been observed. Even if this number is not necessarily well defined, we see that the results are qualitatively the same for any values. Moreover, since we measure fluctuations in the modes frequencies on the order of $\mu$s we can expect that the TLS have a time dynamic on this order of magnitude (a much faster dynamic will average out in our measurement). While some of the TLS within the bandgap of the phononic shield can of course have much longer $\tau_{\text{down}}$~\cite{MacCabe2020}, such TLS would only give a slower frequency shift and would not change the correlation results presented here. We choose a temperature of $T = \SI{1}{K}$ since the results at \SI{800}{mK} are qualitatively the same. However, we note that the Monte-Carlo simulation gives qualitatively identical results for the range of temperature that we can simulate (\SI{0.2}{K} to \SI{4}{K}), with the only difference of the expected decrease in linewidth with higher temperature given by the saturation of the TLS close in frequency to the mechanical modes. We note that the linewidth monotonically decreases with the temperature as in this simple model we do not consider thermal broadening of the mechanical modes. We note that the trend follows a phenomenological decay of $T^{-0.5}$, comparable to previously measured results~\cite{Meenehan2014}, showing that this model captures the interaction between the mechanical modes and the TLS bath quite well. We define the time step $dt$ of the simulation so that all TLS in each time step have a small probability ($<$0.05) of decaying ($P_{\text{i,down}} = (1-\sigma_{z,i}) dt/\tau_{\text{down}}$) or getting excited ($P_{\text{i,up}} = \sigma_{z,i} dt/\tau_{\text{i,up}} $), where $\sigma_{z,i} = 0,1$ is the $i$th-TLS state z projection (0 for ground state, 1 for excited state)~\cite{Berthelot2006}. At each time step, each TLS can decay with probability $P_{\text{i,down}}$ (where $P_{\text{i,down}} <<1$), can get excited with probability $P_{\text{i,up}}$ (where $P_{\text{i,up}} <<1$), or can maintain its state with probability ($1-P_{\text{i,down}}-P_{\text{i,up}}$). We then compute for each time step the total shift in frequency given by the dispersive coupling of the mechanical mode with all the TLSs. For mode A' it will be $\delta_{\text{A'}} = \Sigma_i (g_{i,\text{A'}})^2/\Delta_{i, \text{A'}} \sigma_{z,i}$, summing over all TLSs, where $\Delta_{i, \text{A'}} = \nu_{\text{A'}} - \nu_i$ is the detuning between the $i$th-TLS and mode A'. Similarly, we perform the calculation for B'. We report the time trace for the frequency shift of a single trajectory in Fig.~\ref{Fig:S7_jitter}d. The frequency shifts on short time scale are given by telegraphic noise from the population jumps of the TLS (see Fig.~\ref{Fig:S7_jitter}e). These quantum jumps occur on an effective time scale of $\tau \sim 1/\Gamma_{\text{down}} + 1/\Gamma_{\text{up, av}} \sim\SI{500}{ns}$, where $\Gamma_{\text{up, av}}$ is the average excitation rate. Only using a detection scheme faster than $\tau$ could allow to resolve theses quantum jumps in time. In Fig.~\ref{Fig:S7_jitter}f, we report the histogram of the time traces of all 10 trajectories, with  and without convolution with the detection response. The histogram width of \SI{5}{kHz} is very close to the measured linewidth (re-scaled by a factor 5 to account for the smaller number of TLS considered). Note that this model explains also the differences in linewidth for modes close in frequency. In fact a small number of TLS can more strongly interact with one mode compared to others. In the particular case reported in Fig.~\ref{Fig:S7_jitter}, the mode A' has a much larger linewidth.

\begin{figure}[ht!]
	\centering
	\includegraphics[width = 0.95\linewidth]{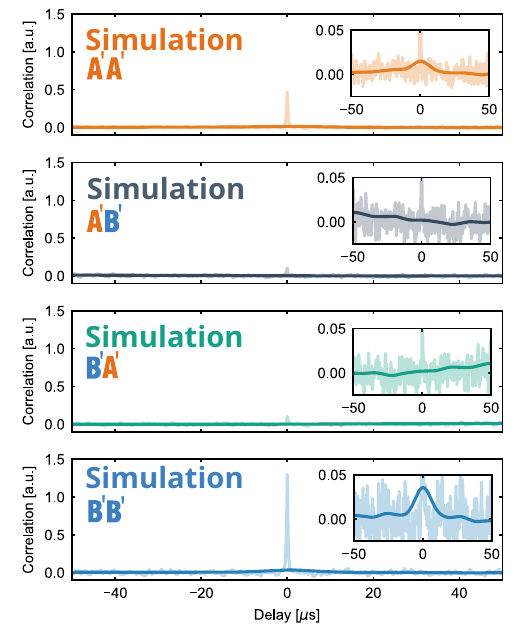}
	\caption{ a) Calculated correlations for the full simulation (10 trajectories) of the signal reported in Fig.~\ref{Fig:S7_jitter} for the possible configuration A'A', A'B', B'A', and B'B' (from top to bottom). Light line for the time trace and dark line for the time trace convoluted with the RSA filter. Note how all time traces show a correlation with a fast decay. However only the autocorrelation shows a peak sensibly higher than the uncorrelated background at 0 delay in the signal convoluted with the RSA filter.}
	\label{Fig:S9_jitter}
\end{figure}

We now calculate the correlations of two time traces of the frequency shift of mode A' and B' in the four possible configurations, A'A', A'B', B'A', and B'B', similar to the main text. To re-normalize we divide the calculated correlations within the same trajectory with the one of different trajectories, since the latter one is completely uncorrelated. We subtract 1 to obtain a value of uncorrelated signal of 0 to be consistent with the main text. In Fig.~\ref{Fig:S9_jitter}, for all the configurations, we show the correlations between the time traces, as well as the correlations between the time traces convoluted with the RSA bandwidth. We see a fast correlation between the time traces, with a time scale of $\tau$, in all four cases. The (small) correlations in the A'B' and B'A' cases is due to the subgroup of TLS which have similar coupling rates and similar detunings for both mechanical modes, causing a correlated shift of the mechanical frequencies. Note that, even if a faster measurement could resolve the (small) cross correlation, the two mechanical modes positions will decohere very fast. For longer times, the correlation is lost and the convoluted signal shows a reduced peak correlation (at zero delay) since it is effectively averaged with an uncorrelated signal (having 1/RBW $>$ $\tau$). In particular, in the cases A'B' and B'A' the convoluted signals show no sensible correlations since the peak value of the correlation between the time traces is extremely small, consistent with our experimental observations.

\end{document}